\RequirePackage{lineno} 
\documentclass[iop]{emulateapj}
\usepackage{graphicx}
\usepackage{epstopdf}
\usepackage{apjfonts}
\usepackage{amsmath}
\usepackage{threeparttable}

\usepackage[bookmarks,breaklinks]{hyperref}
   \hypersetup{
     colorlinks,
     citecolor=blue,
     linkcolor=blue,
    }  
 \defcitealias{2014ApJ...788..120L}{L14}  
 
\begin{document}

%
\shorttitle{Mitigation of Structure Variability in Sgr A*}
\shortauthors{Lu et al.}
\title{IMAGING AN EVENT HORIZON: MITIGATION OF SOURCE VARIABILITY OF SAGITTARIUS A*}
\author{
Ru-Sen \ Lu\altaffilmark{1, 2},
Freek Roelofs\altaffilmark{3},
Vincent L.\ Fish\altaffilmark{1},
Hotaka Shiokawa\altaffilmark{4},
Sheperd S.\ Doeleman\altaffilmark{1, 5},
Charles F.\ Gammie\altaffilmark{4},
Heino Falcke\altaffilmark{3},
Thomas P. Krichbaum\altaffilmark{2},
J. Anton Zensus\altaffilmark{2}
}
\email{rslu@mpifr-bonn.mpg.de}
\altaffiltext{1}{Massachusetts Institute of Technology, Haystack Observatory, Route 40, Westford, MA 01886, USA}
\altaffiltext{2}{Max-Planck-Institut f{\"u}r Radioastronomie, Auf dem H{\"u}gel 69, 53121 Bonn, Germany}
\altaffiltext{3}{Department of Astrophysics, Institute for Mathematics, Astrophysics and Particle Physics, Radboud University, PO Box 9010, 6500 GL Nijmegen, The Netherlands}
\altaffiltext{4}{Astronomy Department, University of Illinois, 1002 West Green Street, Urbana, IL 61801, USA}
\altaffiltext{5}{Harvard-Smithsonian Center for Astrophysics, 60 Garden Street, Cambridge, MA 02138, USA} 
 
\begin{abstract}
The black hole in the center of the Galaxy, associated with the compact source
Sagittarius A* (Sgr A*), is predicted to cast a shadow upon the emission of the
surrounding plasma flow, which encodes the influence of general relativity in
the strong-field regime. The Event Horizon Telescope (EHT) is a Very Long
Baseline Interferometry (VLBI) network with a goal of imaging nearby supermassive black holes (in particular Sgr A* and M87) with angular resolution sufficient to observe strong gravity effects near the event horizon.
General relativistic magnetohydrodynamic
(GRMHD) simulations show that radio emission from Sgr A* exhibits variability on
timescales of minutes, much shorter than the duration of a typical VLBI imaging
experiment, which usually takes several hours. A changing source structure
during the observations, however, violates one of the basic assumptions needed
for aperture synthesis in radio interferometry imaging to work.  By simulating realistic EHT
observations of a model movie of Sgr A*, we demonstrate that an image of the 
average quiescent emission, featuring the characteristic black hole shadow and
photon ring predicted by general relativity, can nonetheless be obtained by
observing over multiple days and subsequent processing of the visibilities
(scaling, averaging, and smoothing) before imaging. Moreover, it is shown that
this procedure can be combined with an existing method to mitigate the effects
of interstellar scattering. Taken together, these techniques allow the black
hole shadow in the Galactic center to be recovered on the reconstructed
image.
\end{abstract}

\section{INTRODUCTION}
\label{introduction}
The compact source at the Galactic center (Sgr A*) makes a very strong case that
it is linked with a $4\times10^6\,M_{\odot}$ supermassive black hole, which due
to its proximity (8\,kpc) spans the largest angle on the sky among all known
black holes~\citep{2001ARA&A..39..309M,2010RvMP...82.3121G,2013CQGra..30x4003F}.
For Sgr A*, one Schwarzschild radius, $R_{sch}$, is  $\sim$0.1 AU that subtends
an angle of $\sim$ 10\,$\mu$as to us. According to general relativity (GR), a
lensed image of the event horizon of Sgr A* (known as the ``black hole shadow'')
will appear \citep{1973blho.conf..215B,1979A&A....75..228L,2000ApJ...528L..13F,2004ApJ...611..996T}
and can now be resolved by the Event Horizon Telescope (EHT), a project to
assemble a VLBI network of millimeter wavelength dishes that aims to resolve
general relativistic signatures in the vicinity of nearby supermassive black
holes~\citep{2008Natur.455...78D,2009astro2010S..68D,2009ApJ...697...45B,2010ApJ...718..446J,2011ApJ...735..110B,2011ApJ...738...38B,2011ApJ...727L..36F,2012Sci...338..355D}.

Horizon scale imaging promises to test basic predictions of GR and improves 
our understanding of the physics responsible for accretion and emission in a
strong gravitational field. In particular, imaging a black hole shadow has been a long-standing goal of black hole astronomy. However, imaging the black hole shadow feature in Sgr A* has been inherently challenged
by two known effects. First, the scattering by interstellar medium blurs the
strong GR features near the black hole. In a recent work, it has been shown that
this effect can be mitigated based on the fact that the scattering is well
understood over the relative range of baseline lengths provided by the
EHT~\citep{2014ApJ...795..134F}.  Second, while the predicted shadow feature is
nearly independent of the spin or orientation of the black hole to within
10\,\%~\citep{1973blho.conf..215B,2004ApJ...611..996T},  the emission region surrounding the black hole depends on the details of the underlying accretion process and is intrinsically time variable primarily
due to the stochastic nature of magnetorotational-instability-driven turbulence and magnetic reconnection in 
the accretion flow. Magnetorotational instability~\citep[MRI,][]{1991ApJ...376..214B,1998RvMP...70....1B} is believed
to be the leading mechanism driving turbulence in accretion disks and develops
on orbital timescales. The timescale for the Keplerian motion at the innermost stable circular orbit around the black hole in Sgr A* ranges from 30 minutes for a non-rotating black hole to 4 minutes for prograde
orbits around a maximally rotating black hole~\citep{2009ApJ...695...59D}. These time scales are much less than the typical duration of a Very Long Baseline Interferometry (VLBI) experiment, which violates one of the basic requirements for VLBI Earth-rotation aperture synthesis imaging. In contrast, the corresponding timescales in the nearby giant elliptical galaxy M87, which has the second largest apparent event horizon, are much larger (a minimal time scale of a few days).

In this paper, we show that the short-time scale structural variability of Sgr\,A* does not prevent construction of time-averaged images that contain distinguishable features like the black hole shadow.
Section~\ref{sect:models} describes the models we employed in this analysis
and Section~\ref{sect:method} details the data simulation, imaging
analysis and quality assessment metrics. Our results on imaging strategy are
presented in Section~\ref{sect:results} and are discussed in
Section~\ref{sect:discussion}. We summarize our conclusions in
Section~\ref{sect:summary}.

\section{GRMHD Simulations of Sgr A*}
\label{sect:models}
We performed time-dependent simulations of black hole accretion using fully conservative 3D GRMHD code
HARM \citep{2003ApJ...589..444G, 2006ApJ...641..626N}. The simulation assumes the flow is radiatively inefficient, as in time-independent, phenomenological models~\citep[e.g.][]{1982Natur.295...17R,1995ApJ...452..710N,2014ARA&A..52..529Y}, which is appropriate for low-luminosity galactic nuclei such as Sgr A*~\citep[see also][]{2013MNRAS.431.2872D}. The simulation starts from a geometrically thick hot disk with its pressure maximum at 24 $GM_{BH}/c^2$ surrounding a black hole with dimensionless spin $a_*\simeq 0.94$. The simulation uses modified spherical-polar coordinates with logarithmically spaced radial grids spanning the range 1.2 to 240 $GM_{BH}/c^2$ and an azimuthal range of $2\pi$ rad. The spatial resolution of the simulation is 260$\times$192$\times$128 cells in radial, poloidal, and azimuthal directions, respectively. The disk is initially seeded with a weak poloidal magnetic field that makes the disk unstable to the MRI. The simulation is run for 14,545 $GM_{BH}/c^3$, which is long enough for saturation of the turbulence to be attained at the pressure maximum (this occurs at about 6,000 $GM_{BH}/c^3$).

\begin{figure}
 \begin{center}
\resizebox{0.75\hsize}{!}{\includegraphics[angle=-90]{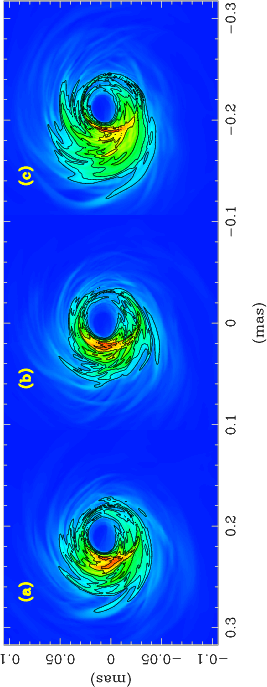}}
 \caption{Sample image frames of Sgr A* taken from the beginning (a), middle (b) and end (c) of the GRMHD simulation (Section~\ref{sect:models}). Contour levels start from 2\,\% of the peak and increase by a factor of 2, which is applicable to all of the subsequent images.}
 \label{fig:frame}
 \end{center}
 \end{figure}

To generate images, we perform general relativistic radiative
transfer on the result of the GRMHD simulation using the \textsc{bothros} ray-tracing code~\citep{2007CQGra..24S.259N}. The main emission source at radio wavelengths is synchrotron radiation from the tenuous magnetized gas. The source function is integrated along geodesics that leads to each pixel of a ``camera''
which is placed 8\,kpc  from the model. A thermal distribution function is
assumed for the electrons. There are several parameters that control the radiative
properties of the model: black hole spin ($a_*$), proton-to-electron temperature
ratio ($r$), and viewing angle ($i$). Here we adopted the model with $a_* \simeq 0.94$, 
$r =3$, and $i=45^{\circ}$ for simulating observations of Sgr A*, which are consistent with existing mm VLBI and spectral measurements, but we also consider simulations with different parameters in the following sections. The length and timescales in the GRMHD model are set by the mass of the black hole, but the density of the accretion flow (equivalently: the accretion rate) is a free parameter. We adjust this free parameter so that the time averaged flux at 230\,GHz after the MRI saturation (t = 6,000--14,545 $GM_{BH}/c^3$) gives 3.4\,Jy, as observed by~\citet{2006PhDT........32M}. The dimensions of the camera frame of the movie are 210 $\times$210 $\mu$as with a resolution of 256$\times$256 pixels.The interval between frames is 221.3 seconds, adding up to a total movie length of 53 hours. Figure~\ref{fig:frame} shows sample images from this simulation.

\section{Method}
\label{sect:method}
\subsection{Data Simulation}
VLBI observations were simulated using the MIT Array Performance Simulator 
(MAPS) software, following~\citet[][hereafter L14]{2014ApJ...788..120L}. Data were simulated at 230\,GHz
with a total bandwidth of 16\,GHz assuming that the model images represent the mean flux over the entire bandwidth. The assumed 16\,GHz bandwidth is consistent with the targeted recording bandwidth of near future EHT observations. As in~\citetalias{2014ApJ...788..120L}, the full EHT 
array was used for simulation, which included the following sites: Submillimeter Array and
James Clerk Maxwell Telescope on Mauna Kea, the Arizona Radio Observatory
Submillimeter Telescope, the Combined Array for Research in Millimeter-wave
Astronomy, the Large Millimeter Telescope (LMT), the Atacama Large
Millimeter/submillimeter Array (ALMA), the Institut de Radioastronomie
Millim\'etrique (IRAM) 30-m telescope on Pico Veleta, the IRAM Plateau de Bure
Interferometer (PdBI), and the South Pole Telescope (SPT). 
For the simulations reported here, telescope elevations were restricted to be above 15 degrees (except 10
degrees for the PdBI), where calibration issues are expected to be reduced.

\begin{figure}
 \begin{center}
\resizebox{0.5\hsize}{!}{\includegraphics{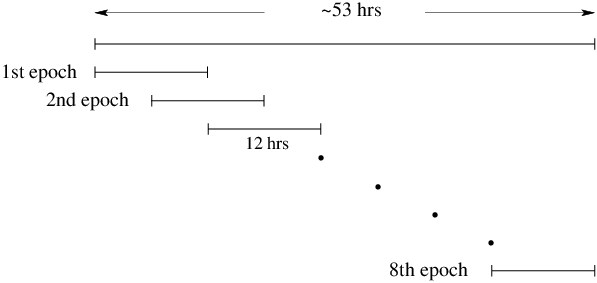}}
 \caption{Sampling of the movie for the case of eight epochs. The {\it uv}-coverage is identical for each epoch except for the last one, where the time coverage is $\sim$1 hr less.}
 \label{fig:diagram}
 \end{center}
 \end{figure}

 As a reference to the underlying quiescent structure, a static image was generated by averaging on a pixel basis all the frames in the GRMHD simulation. Then we generated synthetic VLBI data sets by calculating the complex visibilities and errors on each EHT baseline during a typical night of observing (with $\sim$ 12-hr long total time coverage). These data sets were generated using the static image, and also using the time evolving GRMHD movie with time resolution that matched the movie frame cadence. To simulate multi-epoch data sets, each consecutive block of $\sim$ 12 hrs of the movie (corresponding to 192 frames) was sampled by the array as one epoch with identical {\it uv}-coverage. The duration of the simulation allows 4 epochs of 12-hr long observations without overlap in frames. In order to further increase the number of observing epochs, we also considered a case where the input block of frames for each epoch were sampled with a halfway overlap, leading to a total of 8 epochs. Figure~\ref{fig:diagram} illustrates how the movie was sampled over time in this eight-epoch case.

\subsection{Imaging and quality assessment}
Images were reconstructed with the BiSpectrum Maximum Entropy 
Method~\citep[BSMEM,][]{1994IAUS..158...91B} software, due primarily to its user-friendliness and speediness.
We refer the reader to \citetalias[][and references therein]{2014ApJ...788..120L} for details concerning
imaging reconstruction algorithms. Compared to the widely used
deconvolution-based imaging techniques (e.g., CLEAN), forward imaging
techniques, like the BSMEM, are well suited for mm-VLBI~(\citetalias{2014ApJ...788..120L};\citealt{2014ApJ...795..134F}).

As in~\citetalias{2014ApJ...788..120L}, two image comparison metrics, i.e., mean square error (MSE) and structural dissimilarity (DSSIM) index, were applied to quantify the quality of the reconstructions. 
MSE compares the two images on a pixel-by-pixel basis and it is good for comparing all pixel intensities (Equation 1 in ~\citetalias{2014ApJ...788..120L}). Unlike MSE, DSSIM is derived from the human visual perception metric, structural similarity \citep[SSIM,][]{Wang04} by DSSIM=(1/|SSIM |)-1. SSIM attempts to measure the change in the structural information between the two images by taking into account the change in luminance, contrast, and structure (see Equations 2--6 in \citetalias{2014ApJ...788..120L} for details). In spite of this, these widely used metrics may not be perfectly suited to assessing the reconstruction quality when preserving the visual quality of salient features is crucial. Instead, visual inspection can often do a better job. In any case, future dedicated algorithms being able to detect and characterize specific image features and visual perception experiments appears very important. Here we use the average of all frames of the employed movie as a reference image. For both metrics, lower values indicate better reconstruction quality.

\section{Results}
\label{sect:results}
\begin{figure}
 \begin{center}
\resizebox{0.5\hsize}{!}{\includegraphics[angle=-90]{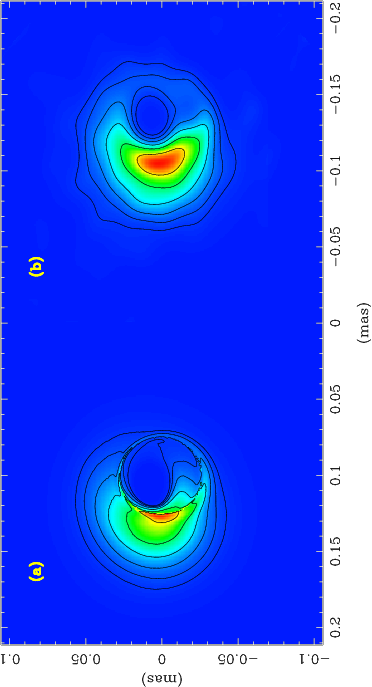}}
 \caption{Image reconstruction of Sgr A*. Average of all movie frames
(a, static structure) and its reconstruction (b, MSE 0.045, DSSIM 0.053). The model image is centered on black hole, while the reconstructed static image is centered on emission centroid.}
 \label{fig:static}
 \end{center}
 \end{figure}

\begin{figure}
 \begin{center}
\resizebox{1.0\hsize}{!}{\includegraphics[angle=-90]{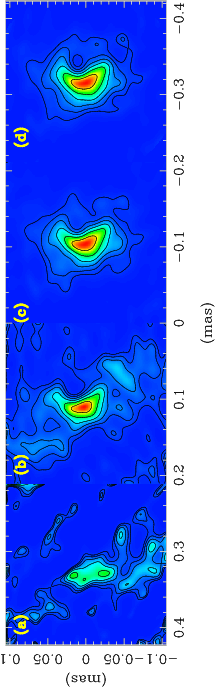}}
\caption{Image reconstruction of Sgr A*. (a) Straightforward reconstruction from a 12-hr observation of the movie simulation (MSE 0.568, DSSIM 0.463), (b) Visibilities from eight epochs are averaged, but not normalized and smoothed (MSE 0.243, DSSIM 0.977).  (c) Visibilities from eight epochs are averaged and normalized, but not smoothed (MSE 0.089, DSSIM 0.192).  (d) Averaging, scaling and smoothing of the complex visibilities obtained from eight epochs were applied (MSE 0.075, DSSIM 0.150).}
 \label{fig:intermediate}
 \end{center}
 \end{figure}

Figure~\ref{fig:static} shows the average image (a) and the
reconstruction of this static structure from a 12-hr observation with the assumed array (b).
The reconstruction produces a high-fidelity image with critical features such as
the photon ring and shadow preserved. However, a straightforward reconstruction of simulated data from a 12-hr movie is very unsuccessful (Figure~\ref{fig:intermediate}, a), indicating that structural variability is a major hurdle to successful horizon-scale imaging of Sgr A*. Averaging of simulated data up to 8-epoch observations improves the reconstruction quality, but noisy features in the reconstruction can make recognition of critical features (e.g., the black hole shadow) difficult in practice (Figure~\ref{fig:intermediate}, b).

\begin{figure*}
\begin{center}
\includegraphics[angle=-90,width=0.495\textwidth,clip]{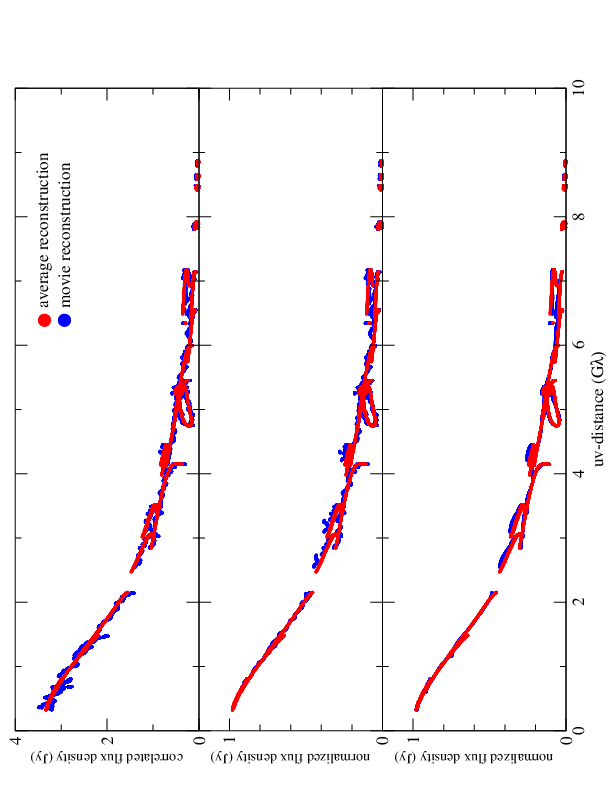}
\includegraphics[angle=-90,width=0.495\textwidth,clip]{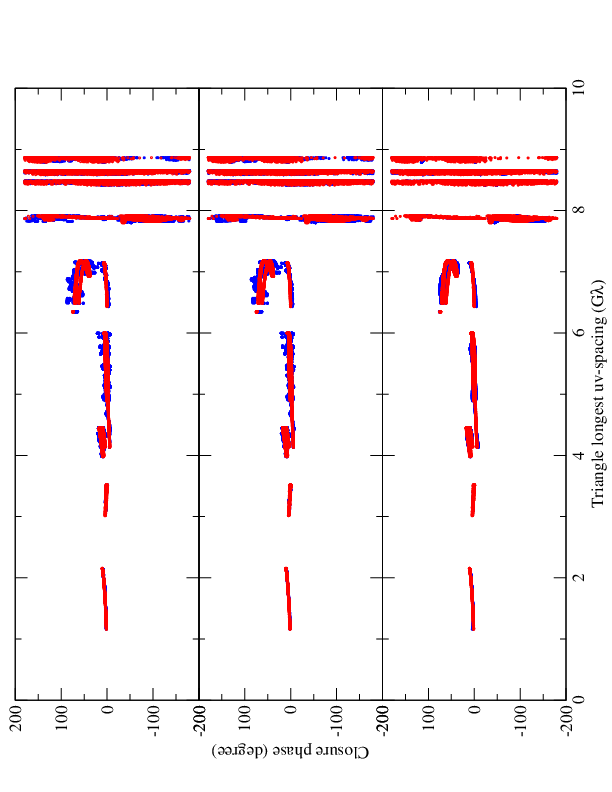}
\caption{Visibility amplitudes (left) and closure phases (right) as a function of baseline length (closure phase plotted against the longest baseline for a given triplet of baselines). In each plot, shown in the upper panel are the visibility amplitudes/closure phases of the static, averaged image (red) and of the averaged visibilities of the movie (blue). Averaged visibilities are then normalized (middle panel), smoothed (lower panel) and are used to calculate closure phases.}
\label{fig:radplot}
\end{center}
\end{figure*}

Figure~\ref{fig:radplot} (left, upper panel) compares the visibility amplitudes of the average image and the averaged amplitudes of the 8 epochs. The average was done in the complex plane for the visibilities with identical {\it u-v} coordinates. The data for the static image look smooth, while for the movie reconstruction there are ``wiggles'' in the visibilities. Closure phases, which are phases of triple products of the complex visibilities around closed triangles~\citep{1958MNRAS.118..276J,1974ApJ...193..293R} and are calculated from the averaged visibilities of the 8 epochs, also show similar effects~(Figure~\ref{fig:radplot}, right, upper panel). This is not surprising because the averaged visibilities are the average of Fourier components of different images. In addition, different movie frames were sampled at different {\it u-v} points. The visibilities at the various {\it u-v} points are thus inconsistent
with both each other and themselves. Furthermore, all visibilities of the static reconstruction contain information about all movie frames, as the image under observation is the average of all frames. On the other hand, each visibility of the movie reconstruction only contains information of the seven or eight frames being averaged. 

In order to reduce this effect, two subsequent data processing steps were applied: scaling and smoothing. The scaling is motivated by the observation that the brightness of pixels in the model images fluctuates in a highly correlated way. That is, there is a component of the variability that can be thought of as scaling the entire image up and down. The total flux information can thus be used to partially remove the variability. The remaining variable component of the structure can be treated as a high-frequency noise on top of the underlying quiescent image and therefore a Fourier smoothing algorithm can be used as a denoising technique. 

For the first step, all visibility amplitudes were normalized by dividing each visibility by the total (zero spacing) flux density of the then observed frame (Figure~\ref{fig:intermediate}, c). In practice, the total flux density can be obtained by continuously observing the source with a connected interferometer. In cases where a mismatch between the measured total flux and the zero-spacing flux density of the horizon-scale structure exists, the scaling factors could possibly be determined by how well the scaling works on short baselines. Scaling significantly reduces the irregularities in the amplitudes on short baselines ($\lesssim$ 2\,G$\lambda$), but does not change the closure phases (Figure~\ref{fig:radplot}, middle panels). After the normalization, significant deviations from the static reconstruction in the amplitudes still exist on baselines longer than $\sim 2$\,G$\lambda$ ( $\gtrsim 4$ G$\lambda$ for closure phases in Figure~\ref{fig:radplot}). A smoothing algorithm was then applied to make the visibilities on baselines longer than $\sim 2$\,G$\lambda$ resemble the reconstruction of the averaged image. This smoothing algorithm is a moving average: each new data point is the average of all old data points within a certain time window, centered on the timestamp of the current data point. This was done in the complex plane for each baseline separately. In order to be able to set a large enough window without losing too much information on large timescales, the data was convolved with a Gaussian weighting function. The smoothing is thus in fact a low pass filter. High frequency structures are averaged out, while longer existing structures are preserved. 

The ``cutoff frequency'' of the filter is determined by the standard deviation
of the Gaussian. If the cutoff frequency is too high, the wiggles in the data
will still be followed closely by the smoothed data. On the other hand, if the
cutoff frequency is too low the smoothing outcome will be a flat line
corresponding to the time-averaged visibility on a particular baseline. A
standard deviation of 100 data points and a window size twice as large gave the
best correspondence to the static reconstruction data. With an integration time of 20s ($\simeq$ $G M_{BH}/c^3$), this standard deviation corresponds to 2000s or 100 $GM_{BH}/c^3$, which corresponds to a few ISCO (Innermost Stable Circular Orbit) rotation periods for a
maximally rotating black hole. The smoothing algorithm thus filters out variability on shorter timescales.
It is worth pointing out that time-averaging of the visibility data smears out the response of a point source located away from the center of the field of view (time smearing effect). Employing an averaging time of 2000 seconds would lead to a fall off of $\sim$10\,\% in the response to the flux $\sim$0.12\,mas from the phase center for a 10000\,km baseline. Since the emission of Sgr A* is known to be very compact~\citep[within $\sim$ 0.04\,mas,][]{2008Natur.455...78D,2011ApJ...727L..36F}, this effect is very small.  

After smoothing, the visibility amplitudes and closure phases of the movie reconstruction show much more overlap with those of the static reconstruction than before (Figure~\ref{fig:radplot}, lower
panels). The resulting reconstruction is shown in Figure~\ref{fig:intermediate} (d), which
shows the same black hole features as the static reconstruction
(Figure~\ref{fig:static}), but slightly less prominent, as indicated by the
increase in MSE and DSSIM.

\begin{figure*}
\begin{center}
\resizebox{1.0\hsize}{!}{\includegraphics[angle=-90]{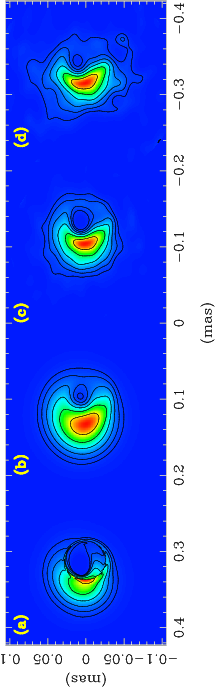}}
\end{center}
\caption{Image reconstruction of Sgr A*. The average of all movie frames (a) is convolved with the scattering kernel (b). The reconstruction of the synthetic visibilities after dividing by the Fourier
transform of the scattering kernel (c) is very close to the original unscattered
static image (MSE 0.045, DSSIM 0.054). The reconstructed image from the scattered
movie (d) using corrected visibilities (averaging, scaling, deblurring and
smoothing) is able to recover the black hole shadow and photon ring (MSE 0.077, DSSIM 0.154).}
\label{fig:scatter}
\end{figure*}

As shown by~\citet{2014ApJ...795..134F}, the effects of interstellar scattering
can be mitigated by correcting the visibilities for the scattering kernel before
imaging. As expected, this technique works well for a static structure
(Figure~\ref{fig:scatter}, a--c). We applied this technique to the movie
reconstruction by convolving all movie frames with the scattering
kernel from~\citet{2006ApJ...648L.127B} before the MAPS simulation.
After the observation, all visibilities were divided by the Fourier transform of
the scattering kernel at their particular {\it u-v} coordinates. Because de-blurring amplifies the thermal noise on long baselines, the degree to which de-blurring works depends on the noise at each telescope, and also on the baseline length. The order in which averaging, scaling and deblurring are performed is irrelevant as they are linear operations, but smoothing was always performed as the last step before
imaging. In Figure~\ref{fig:scatter} (d), the reconstruction from the movie is
able to clearly recover the characteristic signature of the black hole: the photon ring.
From the MSE and DSSIM values, it is clear that scattering and deblurring only marginally decrease the image quality (Figure~\ref{fig:intermediate} (d) and \ref{fig:scatter}).

\begin{figure*}
\begin{center}
\resizebox{1.0\hsize}{!}{\includegraphics[angle=-90]{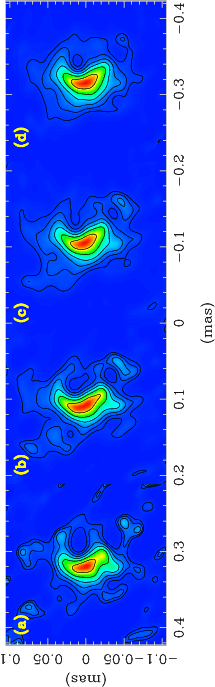}}
\end{center}
\caption{Improvement in reconstructed Sgr A* image quality.
The assumed observing time of the variable structure ranges from 1 to 8 days by a factor of two (from a
to d). The measure of quality metrics of MSE and DSSIM are shown
in Table~\ref{Table:quality}.}
\label{fig:md}
\end{figure*}

The above strategies are indispensable to each other for imaging the 
horizon-scale signatures of Sgr A*. Scaling mainly affects the large-scale
structure (short baselines), whereas smoothing and deblurring mainly affects the small-scale
structure (long baselines). When both techniques are applied to the data, the more
visibilities are averaged, the better the quiescent structure is
approached (Figure~\ref{fig:md} and Table~\ref{Table:quality}). 
In addition,  averaging visibilities will also mitigate refractive noise~\citep[i.e., deviations from ensemble-average scattering,][]{2015arXiv150205722J}. In practice, the source can simply be
observed for multiple days to fulfill this requirement. Since observing the
scattered movie and deblurring the visibilities makes little difference in the
final image, the characteristic shadow and photon ring of the black hole can
indeed be recovered in Sgr A* with the short timescale source variability and
interstellar scattering present.

\begin{table}
\centering
\caption{Quality assessment with MSE (mean square error) and DSSIM (structural
dissimilarity) for images shown in Figure~\ref{fig:md}. \label{Table:quality}}
\begin{tabular}{ccc}
\hline
Duration & MSE & DSSIM\\
\hline
1d & 0.137 & 0.324\\
2d & 0.131 & 0.322\\
4d & 0.107 & 0.276\\
8d & 0.077 & 0.154\\
\hline
\end{tabular}
\end{table}

\section{Discussion}
\label{sect:discussion}
In the current work we have restricted ourselves to the reconstruction of a {\it static} structure out of a GRMHD simulation and have thus used the averaged image as a reference. By averaging in time, however, some of the strong GR effects presented in the GRMHD simulation are smeared out and therefore future horizon-scale imaging should extend the present work to the reconstruction of a {\it variability} image, featuring not only the black hole shadow, but also details of the turbulent accretion flow.

The appearance of the horizon-scale image depends on the black hole properties (mass and spin) and details of the accretion structure and process. Most of these are currently still very uncertain. In this work, we have only considered one time-dependent model to explore the variability mitigation strategy for horizon-scale imaging. Models with different black hole spin, proton-to-electron temperature ratio, and inclination angle may be used to explore the effect of these parameters on the observations, with the knowledge of black hole mass and mass accretion rate. A parameter survey by~\citet{2009ApJ...706..497M} favors $a \sim$ 0.94, $r=3$, and $i=85^{\circ}$, respectively, though the parameters will be model-dependent~\citep[e.g.,][]{2010ApJ...717.1092D,2011ApJ...735..110B}.

\begin{figure*}
\begin{center}
\resizebox{0.5\hsize}{!}{\includegraphics[angle=-90]{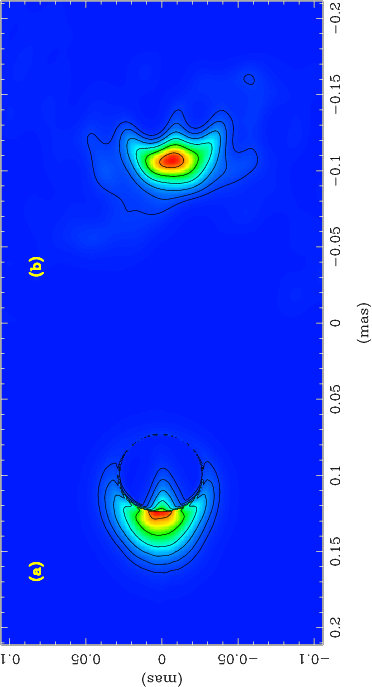}}
\end{center}
\caption{Average of all frames of the model movie with $a \sim$ 0.94, $r=3$, and $i=85^{\circ}$ (a) and a reconstruction of the scattered version of this movie using corrected visibilities from 8-day observations by averaging, scaling, deblurring and smoothing (b, MSE 0.075, DSSIM 0.014).}
\label{fig:inclination}
\end{figure*}

Our procedure, however, is not limited by the uncertainties in the parameter space. As an example, Figure~\ref{fig:inclination} shows a reconstruction of the model movie with $a \sim$ 0.94, $r=3$, and $i=85^{\circ}$ assuming 8-day observations. With a higher inclination angle, the approaching side of the accretion flow becomes brighter and the receding side becomes darker (almost invisible) due to Doppler effects. In this case, the model movie can still be fairly well reconstructed. This suggests that comparison of future EHT observations with simulated observations may tightly constrain model parameters, but to obtain a quantitative comparison new algorithms would have to be developed to detect and characterize features in the image such as the shadow size and position.

Our simulations have implicitly assumed that the visibility amplitude and phase can be measured and calibrated. Accurate amplitude calibration has traditionally been challenging at (sub)millimeter wavelengths, especially when the array is small and limited in sensitivity. However, with more stations being added to the array and planned increases in the data recording rate, the effects of calibration errors should diminish in the near future~\citep[see][for more discussions on the potential improvement on amplitude calibration]{2014ApJ...795..134F}.

On the other hand, fringe phase could also possibly be corrupted by the fluctuations in the atmospheric path lengths. With an array of telescopes of three or more, however, the closure phase, which is inherently robust against station-based phase errors, can be used to retrieve phase information. The fraction of phase information retained by the closure phase monotonically increases with the number of telescopes in the array as $(N-2)/N$, where $N$ is the number of telescopes. Thus, with the anticipated EHT array of eight stations, 75\,\% phase information will be recovered and can be used with measured amplitudes to generate visibilities. In practice, visibility amplitude and closure phase information can be measured in terms of incoherently averaged quantities with well established algorithms to overcome coherence losses~\citep{1995AJ....109.1391R}. Recent EHT observations have shown that both amplitude and closure phase can indeed be successfully measured on Sgr A*~\citep{2008Natur.455...78D,2011ApJ...727L..36F,201xF}. 

\begin{figure}
 \begin{center}
 \includegraphics[width=0.45\textwidth,clip]{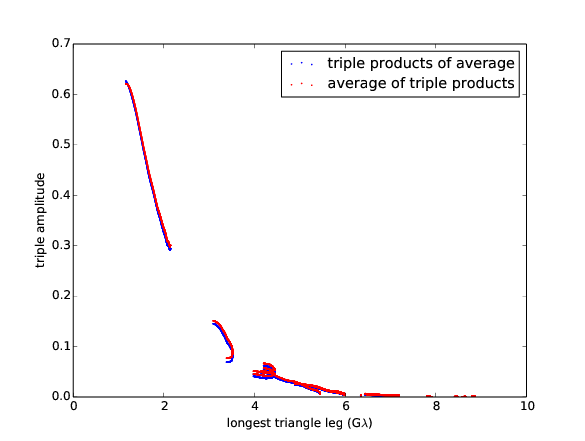}
 \includegraphics[width=0.45\textwidth,clip]{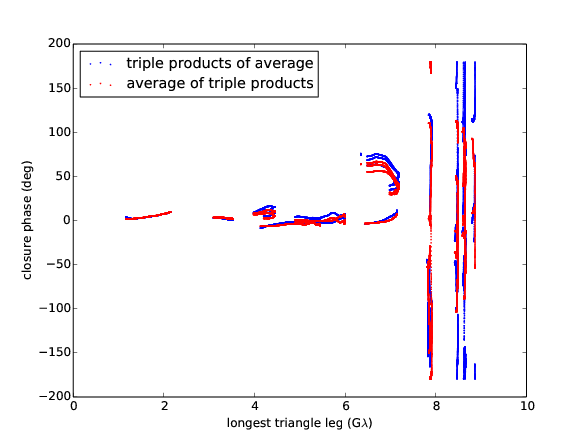}
 \caption{Comparison between averaged triple products (red) and triple products from the averaged visibilities (blue) for the movie simulation. Triple product amplitudes (left) and phases (i.e., closure phases, right) are plotted as a function of the longest baseline length for a given triplet of baselines}.
 \label{fig:averagecphase}
 \end{center}
 \end{figure}

\begin{figure}
 \begin{center}
\resizebox{0.33\hsize}{!}{\includegraphics[angle=-90,clip]{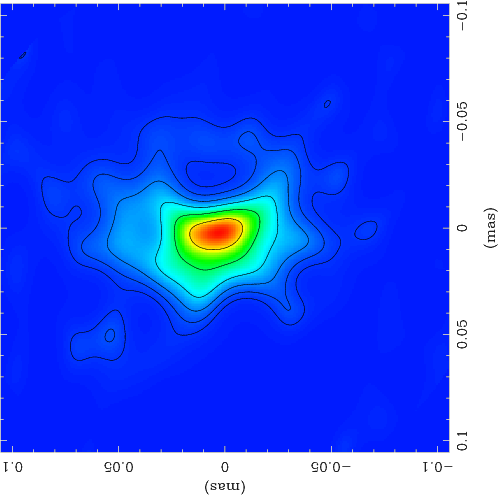}}
 \caption{Reconstructed image of Sgr A* with the same parameters as for the image in Figure~\ref{fig:intermediate} (d), but using averaged triple products (MSE 0.080, DSSIM 0.145)}.
 \label{fig:averageclosure_image}
 \end{center}
 \end{figure}

It is worth pointing out, however, that the triple products from the visibilities averaged over multiple days are not the same as averaging the triple products in the complex plane. Figure~\ref{fig:averagecphase} compares the triple product amplitudes and closure phases for these two cases. The average triple amplitudes and triple amplitudes from the averaged visibilities are similar everywhere. The closure phases, however, are consistent only on small triangles and begin to differ on large triangles (triangle longest leg $\ga$ 4G$\lambda$), where the triple products start rotating rapidly in the complex plane. As a result, using the averaged triple products, the black hole features are less well-preserved (Figure~\ref{fig:averageclosure_image}), indicating sufficient visibility phase information is critical for proper imaging of black hole features.

The source of Sgr A*'s submillimeter variability is not well understood.  Although there will almost certainly
be turbulent variability, as in our GRMHD model, other mechanisms can cause variability on length and timescales
detectable by EHT.  Variability may be caused by orbiting hot spots~\citep[][and references therein]{2009ApJ...695...59D}, jets~\citep{2014A&A...570A...7M}, tilted disks~\citep{2013MNRAS.432.2252D} or episodic particle acceleration.  Our proposed imaging technique may not function equally well in all these cases.  If the variability is very rapid in the uv domain, for example, the width of the Gaussian smoothing kernel will need to be adapted accordingly.  It will therefore be important to test our technique on as large a universe of theoretical models as possible in future studies.

\begin{figure*}
\begin{center}
\resizebox{1.0\hsize}{!}{\includegraphics[angle=-90]{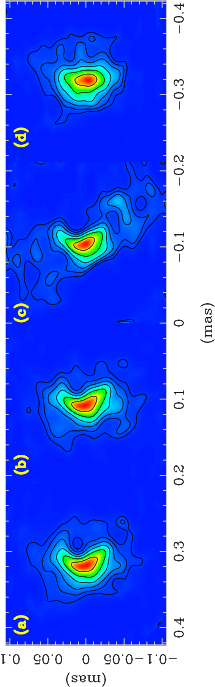}}
\end{center}
\caption{Degradation in reconstructed Sgr A* image quality. Reconstructions are shown in (a) with the full array, in (b) without CARMA, in (c) without Pico Veleta and PdBI, and in (d) without ALMA. For the reconstruction without Pico Veleta and PdBI, each consecutive block of $\sim$ 4 hrs of the movie was sampled as one epoch.}
\label{fig:degradation}
\end{figure*}

\begin{figure*}
\begin{center}
\resizebox{0.75\hsize}{!}{\includegraphics[angle=-90]{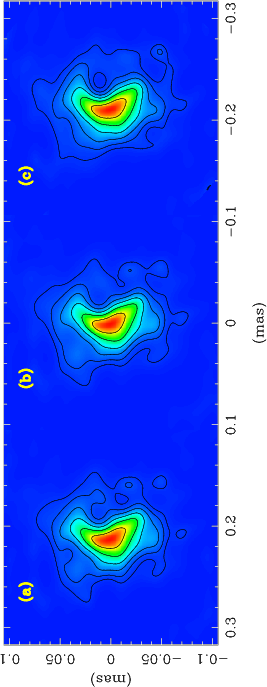}}
\end{center}
\caption{Reconstruction of the movie assuming recording bandwidth of 4, 8, and 16\,GHz. MSE and DSSIM values are (0.090, 0.152), (0.089, 0.176) and (0.077, 0.154), respectively.}
\label{fig:bw}
\end{figure*}

Our ability to image the horizon-scale signatures of the black hole depends on the properties of the observing array. In Figure~\ref{fig:degradation} we show the reconstruction degradation compared to what is obtainable with the full array when a given site is unavailable, e.g., the phased CARMA, Pico Veleta and PdBI, or phased ALMA. A visual inspection indicates that the most severe degradation happens when the phased ALMA is missing (panel d). 
This is because all the longest and most sensitive baselines are provided by the phased ALMA. 
However, both the MSE and DSSIM statistics  (Table~\ref{Table:degradation}) do not confirm the same assessment as perceived by human observers, indicating that these pixel-based metrics provide little understanding on how the morphology of black hole features differs from image to image. Future development of feature-based metrics (i.e., metrics that characterize the morphological properties of black hole features) can potentially provide a more unbiased way for black hole image comparison. Figure~\ref{fig:bw} shows the recording bandwidth impact on the reconstruction fidelity from 4 to 16\,GHz by powers of two. As suggested by the MSE, the image quality gets better with wider recording bandwidths (i.e., higher sensitivity), although the DSSIM does not follow the same trend. Since the bandwidth ($\Delta\nu$) and integration time ($t$) equivalently improve the signal-to-noise ratio of a coherently integrated signal as $\sqrt{\Delta\nu t}$, it is preferable to record data with maximum bandwidth, as the atmospheric coherence time is usually short ($\sim$10s).

\begin{table}
\centering
\caption{Quality assessment for the reconstructed image when a given site is unavailable~\label{Table:degradation}.}
\begin{tabular}{ccc}
\hline
Missing site&MSE&DSSIM\\
\hline
None          &0.077&  0.154\\
CARMA      &0.099&  0.186\\
PV, PdBI    &0.139&  0.522\\
ALMA         &0.092& 0.102\\
\hline
\end{tabular}
\end{table}
 
\section{SUMMARY}
\label{sect:summary}
We have shown that the variability of Sgr A* at 1.3 mm can be significantly mitigated
and the general relativistic black hole features, such as the shadow and
photon ring, predicted by a GRMHD model movie of Sgr A*, can in principle be 
imaged by the EHT. To get a high-quality image, it is essential to observe Sgr A* for multiple days and to 
average visibilities in the complex plane before imaging. Normalizing the
visibilities with respect to the zero-spacing flux density, which can be 
measured in practice, is an important tool to obtain high image quality,
especially for large-scale structures. Applying a smoothing algorithm and
increasing the observation time will further increase the image quality. 
If the properties of the scattering kernel are well known, the reconstructions
can be corrected for interstellar scattering. 

The inclusion of phased ALMA in future observations and recording with wide bandwidth will be critical for imaging the black hole shadow. In order to detect and characterize the black hole shadow features, development of dedicated algorithms is needed in near future studies. Given the currently limited understanding of the origin of flaring structures in Sgr A*, it is also important to explore a wider range of time-dependent source models to improve the capabilities of the proposed imaging techniques.

\acknowledgments
We thank the anonymous referee for comments and suggestions that improved the quality of the paper.
High frequency VLBI work at MIT Haystack Observatory and Smithsonian Astrophysical Observatory is supported by grants from the National Science Foundation (AST-1310896, AST-1211539, AST-1207704, AST-1440254, AST-1337663) and by grant GBMF-3561 from the Gordon and Betty Moore foundation.

 \end{document}